\documentclass[
reprint,
superscriptaddress,
nofootinbib,
amsmath,amssymb, preprintnumbers,
aps,
]{revtex4-2}

\usepackage{caption}
\usepackage{graphicx}
\usepackage{verbatim}
\usepackage{dcolumn}
\usepackage{bm}
\usepackage{subcaption}
\usepackage{hyperref}
\usepackage{stackengine}
 \hypersetup{
     colorlinks= true,
     linkcolor = red,
     citecolor = red,      
     }
\usepackage{tikz-feynman,contour}

\begin{document}

\preprint{QMUL-PH-25-21}
\title{Analytic structure of the high-energy gravitational amplitude: \\ multi-H diagrams and classical 5PM logarithms}

\author{Francesco Alessio}\email{Francesco.Alessio@lnf.infn.it} 
\affiliation{INFN, Laboratori Nazionali di Frascati, 00044 Frascati (RM), Italy }%
\author{Vittorio Del Duca}\email{delduca@lnf.infn.it} 
\affiliation{INFN, Laboratori Nazionali di Frascati, 00044 Frascati (RM), Italy }%
\author{Riccardo Gonzo}\email{r.gonzo@qmul.ac.uk} 
\affiliation{Higgs Centre for Theoretical Physics, School of Physics and Astronomy, 
University of Edinburgh, EH9 3FD, UK}%
\affiliation{Centre for Theoretical Physics, Department of Physics and Astronomy,
Queen Mary University of London, London E1 4NS, United Kingdom}
\author{Emanuele Rosi}\email{emanuele.rosi@uniroma1.it} 
\affiliation{INFN, Laboratori Nazionali di Frascati, 00044 Frascati (RM), Italy }%
\affiliation{Dipartimento di Fisica, Sapienza Università di Roma, Piazzale Aldo Moro 5, 00185, Roma, Italy}%
\author{Ira Z. Rothstein}\email{izr@andrew.cmu.edu} 
\affiliation{Department of Physics, Carnegie Mellon University, Pittsburgh, PA 15213, USA}%
\author{Michael Saavedra}\email{msaavedra@physics.ucla.edu} 
\affiliation{Department of Physics, Carnegie Mellon University, Pittsburgh, PA 15213, USA}%
\affiliation{Mani L. Bhaumik Institute for Theoretical Physics,
 University of California at Los Angeles, Los Angeles, CA 90095, USA}%

\begin{abstract}
We investigate the high-energy, small-angle limit of two-body gravitational scattering. Using power counting arguments and dispersion relations in an effective field theory for the Regge regime, we derive the general loop expansion that determines how the leading Regge logarithms and their complex structure arise as a power series in $t/s$. Focusing on the tower of multi-H diagrams that govern the leading logarithmic behavior, we compute the leading double logarithm at four loops (5PM) using both effective field theory methods and the multi-Regge expansion, finding complete agreement. Finally, using the aforementioned dispersion relations, we extract the single logarithmic contribution to the imaginary part of the eikonal phase at 5PM in the Regge limit.
\end{abstract}

\maketitle

\section{Motivation and introduction}\label{sec:intro}

\!\!\!The Regge limit of QCD has a long and rich history\footnote{See~\cite{DelDuca:1995hf,DelDuca:2022skz,Raj:2025hse} for recent reviews on this topic.}.
In this limit $s\gg|t|$, radiative corrections to $2\!\to\!2$ amplitudes exhibit iterative evolution in the rapidity $y\simeq\log(s/|t|)$. This evolution can be formulated either in the $t$ channel (reggeized exchanges) or in the $s$ channel (rapidity evolution of projectile/target degrees of freedom), and both viewpoints are needed for unitarity~\cite{Lipatov:1989bs}. In practice, only a few iterative structures have been studied in detail. The first is the Balitsky-Fadin-Kuraev-Lipatov (BFKL) equation~\cite{Lipatov:1976zz,Kuraev:1976ge,Kuraev:1977fs,Balitsky:1978ic}, which describes one- and two reggeized-gluon ladders in the $t$-channel and has been further generalized to multi-Reggeon states in the planar limit~\cite{Lipatov:1993qn,Lipatov:2009nt,Lipatov:1993yb,Faddeev:1994zg}. Interestingly, its dynamics can be formulated as a rapidity renormalization group equation (RRGE)~\cite{Chiu:2011qc,Chiu:2012ir}, with the BFKL kernel playing the role of the anomalous dimension~\cite{Rothstein:2016bsq}. The second is the Balitsky-JIMWLK hierarchy~\cite{Balitsky:1995ub,Kovchegov:1999yj,Iancu:2001ad}, which governs the rapidity evolution on the $s$-channel side and provides a complementary, unitarity description at high energies.

In this paper, we will be interested in the behavior of gravity in the Regge limit, for which much less is known. The leading high-energy contributions to the elastic amplitude are governed by the eikonal phase, which arises from multiple soft graviton exchanges in the $t$-channel, see~\cite{DiVecchia:2023frv} for a review. By contrast, the graviton Regge trajectory is suppressed by $t/s$~\cite{Lipatov:1982vv,Bartels:2012ra,Melville:2013qca,Rothstein:2024nlq} and thus represents a quantum correction. Similarly, soft exchanges between Glauber gravitons beyond the single-emission case arise only quantum-mechanically: these correspond to unitarity cuts of $2 \to 2+n$ amplitudes with $n\ge 2$~\cite{Britto:2021pud,Cristofoli:2021jas}, which generate the gravitational BFKL evolution~\cite{Lipatov:1982vv,Lipatov:1982it,Rothstein:2024nlq}. The single soft emission yields the well-known ${\cal O}\left( G_N^3 s^3 \log(s/|t|) \right)$ contribution to the imaginary part of the two-loop amplitude -- the {\it H-diagram} ~\cite{Amati:1990xe}-- obtained from the three-particle unitarity cut in multi-Regge kinematics (MRK).

The leading Regge logarithms at higher loop orders arise by iterating the same structure that appears in the H-diagram: each new rung corresponds to inserting an additional three-particle unitarity cut in MRK, with a single on-shell graviton connecting pair of different rungs. This construction defines a {\it multi-H generalisation } of the H~diagram~\cite{Amati:1993tb,Amati:2007ak,Ciafaloni:2015xsr}, which contribute at $(2N+3)$ Post-Minkowskian (PM) order and provides a correction of ${\cal O}\left( (G_N^2 s \log(s/|t|))^N \right)$ to the H~diagram appearing at 3PM order~\cite{Amati:1990xe}. These classical logs are particularly relevant for the gravitational wave community, especially given the ongoing interest in targeting the 5PM order~\cite{Driesse:2024xad,Driesse:2024feo,Dlapa:2025biy,Bern:2025zno}. Here, we take this opportunity to compute the double logarithm appearing in the real part, as well as the single logarithm in the imaginary part, of the amplitude at 5PM order, using both effective field theory (EFT) and Regge techniques~\cite{Rothstein:2024nlq,Alessio:2025Regge}.

The S-matrix viewpoint suggests that the high-energy, fixed-$t$ amplitude is constrained by analyticity, crossing, and unitarity, yielding subtracted dispersion relations that organize the tower of $\log(s/|t|)$ terms~\cite{Collins:1977jy}. The behavior of the gravitational amplitude in the forward limit is well captured by the familiar eikonal formula (see~\cite{DiVecchia:2021bdo} for a review), which reproduces the graviton pole~\cite{Haring:2024wyz} and saturates the relevant spin-2 dispersive sum rules~\cite{Caron-Huot:2021rmr,Caron-Huot:2021enk,Chang:2025cxc}, suggesting that a twice-subtracted fixed-$t$ dispersion relation holds for the $2\to 2$ amplitude~\cite{Caron-Huot:2022jli,Haring:2022cyf}. This has indeed been applied at 3PM in the Regge limit~\cite{Amati:1990xe,DiVecchia:2020ymx}, which motivates us to revisit and extend such arguments to higher loops in the ultra-relativistic regime.

This paper is organised as follows. We first establish the relevant power counting for the EFT of forward scattering in gravity in Sec.~\ref{sec:high-energy}, emphasizing the distinction between classical and quantum contributions. In Sec.~\ref{sec:EFT} we then review the application of this formalism to amplitudes and set up the relevant RRGE at leading-log accuracy, making contact with the gravitational BFKL kernel. In Sec.~\ref{sec:multiH} we construct the multi-H series from both the EFT perspective and the Regge-theory approach in MRK, demonstrating the complete agreement between these approaches and computing explicitly the single and double-H diagrams in $D$ dimensions. Finally, in Sec.~\ref{sec:disp_rel} we derive new dispersion relations for gravitational amplitudes as an expansion in powers of $t/s$ obtaining, under the assumption of eikonalization, the leading logarithmic terms in the real and imaginary parts of the classical eikonal phase at 5PM order.

\textit{Conventions---}
We use the mostly-minus metric signature $\eta=(+,-,-,-)$. Gravitational coupling and Newton's constant are related by $\kappa=\sqrt{8\pi G_N}$. Momentum-space integrals are defined with the measure $\hat{\mathrm{d}}^Dp = \mathrm{d}^D p/(2\pi)^D$ in $D=4-2\epsilon$ dimensionally regularized dimensions. We adopt the shorthand notation $\hat{\delta}^{(D)}(\,\cdot\,)\!=\!(2\pi)^D{\delta}^{(D)}(\,\cdot\,)$ and $\delta^+(k^2)\!=\!\Theta(k^0) \delta(k^2)$ and employ light-cone coordinates $(p^+,p^-,p_{\perp})$, with $p_{\perp}$ a transverse vector in $d\!=\!D\!-\!2$ dimensions. Finally, we define the symbol $\int_{{q}_{\perp}}\!=\!\int \hat{\mathrm{d}}^d q_{\perp}$ for the integration in transverse space.

\section{Power Counting in Gravity and When it Fails}
\label{sec:high-energy}

The effective field theory describing forward scattering was developed for QCD in~\cite{Rothstein:2016bsq} and extended to gravity in~\cite{Rothstein:2024nlq}.  These constructions generalize the framework of soft-collinear effective theory (SCET), originally formulated for hard scattering in QCD~\cite{Bauer:2000ew,Bauer:2001yt,Bauer:2000yr} and subsequently adapted to gravity~\cite{Beneke:2012xa,Okui:2017all,Beneke:2021aip}. The forward scattering theory differs from its hard scattering counterpart through the presence of a Glauber (off-shell) mode, which dominates the interaction in the Regge limit.  Moreover, the power counting is especially distinct in the gravitational case due to the mass dimension of the coupling. 

We consider Einstein's gravity with the hierarchy
\begin{equation}
   s \gg \frac{1}{G_N} \gg |t|\,,
\end{equation}
where $s$ and $t$ are the Mandelstam invariants of the two-body scattering amplitude $\mathcal{M}_{2 \to 2}$. The corresponding EFT is organized in terms of the expansion parameters
\begin{equation}
\alpha_Q \equiv G_N t, 
\qquad 
\alpha_C \equiv G_N^2 s t,
\qquad 
\lambda \equiv \frac{t}{s},
\end{equation}
and we work only to leading order in~$\lambda$\,\footnote{While $t/s$ is not independent of the other two parameters, it is convenient to treat it as a separate expansion parameter.} 
but to all orders in~$\alpha_C$ and~$\alpha_Q$, corresponding to the ``classical'' and ``quantum'' couplings, respectively. As shown in App.~\ref{app:massless-appendix}, this is equivalent to working at leading logarithmic accuracy in the rapidity logarithms $\log(s/|t|)$; correspondingly, we neglect subleading-power corrections in $\lambda$ as well as hard-region contributions, such as double logarithms, that lie outside the forward/Glauber EFT.

The difficulty with the gravitational perturbative series arises from the fact that the effective gravitational charge grows with energy, scaling as $G_N s = \alpha_C / \alpha_Q$. This means that at high energy, the expansion cannot be uniformly controlled unless we restrict to observables that are insensitive to short-distance (contact) interactions. Such observables correspond to scattering between well-separated particles, described by localized wave packets with large angular momentum~\cite{Kosower:2018adc}. 
Even within this regime, however, local UV-sensitive effects can generate nonlocal contributions at higher loops. Calculational control is regained if the amplitude eikonalizes in impact-parameter space~\cite{DiVecchia:2023frv}, where it takes the form
\begin{equation}
    i \int_{q_\perp} e^{i q\cdot b} \frac{\mathcal{M}_{2 \to 2}(s, t = q^2)}{2s}
    = \bigl(1 + \Delta_Q \bigr) e^{2 i \delta_{\text{Cl}}} - 1,
    \label{Amp Eik}
\end{equation}
with $\delta_{\text{Cl}}$ and $\Delta_Q$ representing the classical and quantum contributions, respectively, expanded as
\begin{align}
    \delta_{\text{Cl}} &= G_N s \sum_{j=0}^{\infty} \alpha_C^{j}\, \delta_{\text{Cl}}^{(2j)}(b), \nonumber\\
    \Delta_Q &= \sum_{n,\,k=0}^{\infty} \alpha_Q^{n} \alpha_C^{k + 1}\, \Delta_Q^{(n,k)}(b)\,.
\end{align}
For massive scattering, with $m^2 \gg |t|$, eikonalization is known to hold~\cite{Levy:1969cr,Akhoury:2013yua,Du:2024rkf}, whereas in the strictly massless case evidence from $\mathcal{N}=8$ supergravity suggests that it can fail at subleading-logarithmic accuracy starting at 4PM~\cite{DiVecchia:2019kta}.
Although in the EFT we will work in the massless case, our Regge theory approach originates from the massive regime and smoothly extends to the massless limit at leading order in~$\lambda$. 
Consequently, at leading logarithmic accuracy our target calculations will not be affected by any possible eikonalization breakdown.


\section{EFT of Forward scattering for Gravity}
\label{sec:EFT}

Within the EFT framework, it can be shown~\cite{Rothstein:2016bsq,Gao:2024qsg,Rothstein:2024fpx} that the $2\!\to\!2$ amplitude factorizes, at leading power, into a convolution in transverse momentum space
 \begin{equation}
 \mathcal{M}_{2\rightarrow 2}  
    = i \sum_{M,N} 
    J_{(M)} \!\otimes\! S_{(M,N)} \!\otimes\! \bar{J}_{(N)}\,,
    \label{eq:Regge_factorization}
 \end{equation}
where $J_{(M)}$, $\bar{J}_{(N)}$ and $S_{(N,M)}$ depend on the collinear and soft degrees of freedom respectively. In general, the soft function can mix sectors with different numbers of Glaubers, but in gravity at the order of interest $S_{(M,N)} \propto \delta_{MN}\,S_{(M)}$~\cite{Rothstein:2024nlq}, so the sum reduces to a single index.

The dependence on the large ratio $s/|t|$ enters through logarithms of the rapidity renormalization scale $\nu$ (the analog of $\mu$ in dimensional regularization), which separate the collinear and soft modes. The resummation of these rapidity logarithms is governed by the RRGE, which can be implemented by evolving either the collinear functions down to $t$ or, equivalently, the soft function up to $s$; here we adopt the latter viewpoint. Consequently, all leading logarithms can be obtained by studying the evolution of $S_{(N)}$ in isolation, while subleading logarithms may receive collinear or finite-mass corrections. In particular, the leading-log structure is universal and identical in the massive and massless theories.

Each matrix element satisfies a RRGE~\cite{Chiu:2011qc,Chiu:2012ir}, with the soft function  in particular satisfying
 \begin{align}
 \nu \frac{d}{d\nu} S_{(M)} 
 &= -\,\gamma_{(M)} \otimes S_{(M)} 
     - S_{(M)} \otimes \gamma_{(M)}\,,
 \label{RRGE}
 \end{align}
where $\gamma_{(M)}$ is the anomalous dimension of the $M$-Glauber soft matrix element.  At leading-logarithmic order (i.e. for the one loop anomalous dimension) it takes the form
\begin{align}
\label{eq:anomalous_dim}
\gamma_{(M)} \sim &\sum_{i}\omega_G(q_i)I_{\perp(M-1)}  \nonumber \\
& +\sum_{\text{pairs}\, i,j}\mathcal{K}^{\text{GR}}(q_{i},q_j;q)\,I_{\perp(M-2)}\,.
\end{align}
In the above, $\omega_G$ is the graviton Regge trajectory and $\mathcal{K}^{\text{GR}}$ is the gravitational BFKL kernel~\cite{Lipatov:1982vv,Lipatov:1982it}
\begin{equation}
 \mathcal{K}^{\mathrm{GR}}(q_1,q_2;\,q)
 \;\equiv\;\mathcal{K}^{\mathrm{GR}}(q_1,q_2;\,q-q_1,\,q-q_2),
\end{equation}
where the four–argument form reads, 
 \begin{align}
 &\mathcal{K}^{\mathrm{GR}}(q_1,q_2;\,q_3,q_4) = \frac{4}{(q_{1}-q_{2})_{\perp}^4}\Big[q_{1\perp}^2 q_{2\perp}^2 q_{3\perp}^2 q_{4\perp}^2 \\
 &\qquad - q_{3\perp}^2 q_{4\perp}^2 (q_{1\perp}\!\cdot\! q_{2\perp})^2 - q_{1\perp}^2 q_{2\perp}^2 (q_{3\perp}\!\cdot\! q_{4\perp})^2\Big]\nonumber\\
 &\qquad  +\Big[(q_{1\perp}+q_{3\perp})\!\cdot\!(q_{2\perp}+q_{4\perp})
 -\frac{q_{2\perp}^2 q_{3\perp}^2+q_{1\perp}^2 q_{4\perp}^2}{(q_{1}-q_{2})_{\perp}^2}\Big]^2 \,. \nonumber 
 \end{align}
The first sum in~\eqref{eq:anomalous_dim} corresponds to the insertion of the Regge trajectory on any individual Glauber exchange, while the second term runs over all pairs of Glauber rungs connected by a BFKL kernel. The operator $I_{\perp(M)}$ acts as the identity on the convolution space for $M$ Glauber exchanges.  High-energy logarithms in the amplitude can then be predicted  by evolving in~$\nu$ from the soft rapidity scale $\sqrt{-t}$ up to the collinear rapidity scale $\sqrt{s}$.

At tree level, the soft and collinear matrix elements scale as $J_{(M)} \otimes S_{(M)} \otimes \bar{J}_{(M)} \sim (G_N s)^M s/t $ while the one-loop anomalous dimension scales as $\gamma_{(M)} \sim G_N t$. Taken together, the RRGE~\eqref{RRGE} then predicts a classical contribution from $S_{(M)}$ to the amplitude at $(2M-1)$-PM order of the form $\log^{M-1}(s)$~\cite{Rothstein:2024nlq}. In general, this term will contain both genuinely classical contributions as well as quantum iterations, the latter involving insertions of the graviton Regge trajectory as well as higher-order BFKL-type ladders with two or more graviton exchanges between the same pair of Glauber rungs. When solving Eq.~\eqref{RRGE}, these non-classical pieces can be directly discarded, leaving only the contributions in which each Glauber rung exchanges a graviton with at least one other rung. These configurations are exactly the classical $s$-channel multi-H diagrams~\cite{Amati:2007ak,Ciafaloni:2015xsr}.

\section{Multi-H diagrams from amplitudes and EFT tools}
\label{sec:multiH}

In this section, we will calculate the leading logarithmic contributions to the classical amplitude, recovering the single $\log(s/|t|)$ at 3PM and then computing the double $\log^2(s/|t|)$ at 5PM. To do so, we will use both the EFT approach and Regge-theory techniques, demonstrating the complete agreement between the two approaches.

\subsection{The EFT approach}

In the EFT framework, the leading rapidity logarithms arise from the evolution of the soft function under the RRGE~\eqref{RRGE}, where each iteration inserts an additional soft graviton exchange between distinct Glauber lines and generates one power of $\log s$. 

The normalization of the soft function and its anomalous dimension is fixed by the factorization formula~\eqref{eq:Regge_factorization},  which can be made explicit as
\begin{align}
      \label{eq:ampFactorized}
\mathcal{M}_{2\rightarrow2} & = i \sum_{M}
   \int_{\!\perp(M\times M)} 
   J_{(M)}(\{l_{i\perp}\})\\
   &\qquad \qquad \times  S_{(M)}(\{l_{i\perp}\};\{l'_{i\perp}\})
   \bar{J}_{(M)}(\{l'_{i\perp}\})\,,\nonumber 
\end{align}
where the transverse convolution measure is
\begin{align}
\label{eq:convN}
&\int_{\!\perp(M\times M)}  \equiv \int_{\!\perp(M)}  \int_{\!\perp(M)}\,, \nonumber \\
  \int_{\!\perp(M)} =
  &\frac{(-i)^{M}}{M!}
  \Bigg[\prod_{i=1}^{M}\int_{l_{i\perp}}
  \frac{1}{l_{i\perp}^2}\Bigg]\!\hat{\delta}^{(d)}\Big(\sum_i l_{i\perp}-q_\perp\Big).
\end{align}
While we do not know how to solve the RRGE equation~\eqref{RRGE} in general, the leading logarithm can be computed systematically by iterating such equation. The classical (Regge-cut) part of~$\gamma_{(M+1)}$ in \eqref{eq:anomalous_dim} corresponds to the exchange of a single soft graviton connecting any pair of the $(M+1)$ Glauber lines, and is given by~\cite{Rothstein:2024nlq}
\begin{align}
\gamma^{\text{Cl}}_{(M+1)}
 &= -\,i^{\,M+1}(M+1)!
    \sum_{i<j}
    \frac{\kappa^2}{8\pi}\,
    \mathcal{K}^{\mathrm{GR}}(k_{i\perp},k_{j\perp};\ell_{i\perp},\ell_{j\perp})\, \nonumber \\
    &\qquad \qquad \quad \,\times \prod_{m\neq i,j}\!
    \ell_{m\perp}^2\,
    \hat{\delta}^{(d)}(\ell_{m\perp}-k_{m\perp}).
\label{eq:gamma_classical}
\end{align}
The tree-level value of the soft function is
\begin{align}
&S_{(M)}^{(0)}\!\left(\{l_{i\perp}\};\{l'_{i\perp}\}\right)
= 2\,i^{M}\,M!\;
\Bigg[\prod_{a=1}^{M} \big(l'_{a\perp}\big)^{2}\Bigg] \nonumber \\
&\qquad \qquad \qquad \qquad \times \Bigg[\prod_{n=1}^{M-1} \hat{\delta}^{(d)}\!\big(l_{n\perp}-l'_{n\perp}\big)\Bigg].
\label{eq:SN_tree}
\end{align}

A single iteration of the RRGE~\eqref{RRGE} acting on the two--Glauber soft function $S^{(0)}_{(2)}$ reproduces the single H diagram~\cite{Rothstein:2024nlq}, giving the $\mathcal{O}(G_N^3 s^3 \log s)$ contribution to the amplitude at 3PM order.  The next iteration involves the three--Glauber soft function $S^{(0)}_{(3)}$, whose running yields the $\mathcal{O}(G_N^5 s^4 \log^2 s)$ double-H contribution at 5PM order. We will discuss those more in details in the next sections, using the complementary Regge-theory perspective.

\subsection{The Regge theory approach}

The multi-H diagrams obtained from the RRG can be equivalently constructed in the traditional amplitude approach in MRK~\cite{Alessio:2025Regge}. One starts from the $M$-loop ladder diagram, which contains $M+1$ vertical graviton propagators and thus carries the usual eikonal factor $1/(M+1)!$ from the symmetrization of identical exchanges. The $M$ horizontal soft gravitons are then inserted by connecting distinct pairs of vertical lines, with no repeated pairs in order to exclude vertical graviton loops, which are absent in the classical limit.  The number of available pairs is $N_{\rm pairs}=\binom{M+1}{2}$, so the number of topologically distinct multi-H topologies is 
\begin{align}
C\!\left(N_{\rm pairs},\,M\right)=\frac{N_{\rm pairs}!}{(N_{\rm pairs}-M)!}\,.
\label{eq:counting_top}
\end{align}
In addition, the soft graviton phase space with strong rapidity ordering $y_1 \gg y_2 \gg \dots \gg y_{M}$ gives
\begin{align}
\label{eq:log}
\frac{\log^{M} (s)}{M!} 
&= \int \mathrm{d}y_1 \cdots \mathrm{d}y_{M} \;
\Theta(y_1 > y_2 > \dots > y_{M})\,,
\end{align}
where the factor $1/M!$ arises from cutting $M$ identical soft gravitons, i.e. from restricting to the ordered rapidity domain. In \eqref{eq:log} and throughout, the argument of the logarithm is understood to be $s/\mu^2$, with $\mu^2$ a scale of order $q_\perp^2$. 
In the Regge limit, these soft graviton exchanges are placed on shell and thus correspond to genuine multi–particle intermediate states. Their effect is to generate Regge cuts, which may be understood as arising from overlapping discontinuities in multi--particle kinematics. We now evaluate explicitly the contributions from the single and double-H diagrams.

\paragraph*{Single H diagram:} Here we revisit the computation of the H diagram originally carried out in~\cite{Amati:1990xe}, where it was used -- together with real analyticity and crossing symmetry -- to extract the full 3PM eikonal in the ultra-relativistic limit. More recently, this analysis was extended to the massive case~\cite{DiVecchia:2020ymx}, and employed as a tool to obtain the 3PM eikonal in $\mathcal{N}=8$ supergravity~\cite{DiVecchia:2021bdo} (see also~\cite{Bern:2019nnu,Damour:2019lcq,Parra-Martinez:2020dzs,Kalin:2020fhe,Damour:2020tta,Herrmann:2021tct,Brandhuber:2021eyq,Bern:2020gjj,Cheung:2020gyp,Bjerrum-Bohr:2021din,Bjerrum-Bohr:2021vuf} for related studies at 3PM order).

\begin{figure}[t]
  \centering
  \begin{tikzpicture}[scale=1.0,baseline=(b.center)]
    \begin{feynman}
      \vertex (in1)  at (-2,  1.3);
      \vertex (v1)   at (-1,  1.3);
      \vertex (v2)   at ( 1,  1.3);
      \vertex (out1) at ( 2,  1.3);
      \vertex (in2)  at (-2, -1.3);
      \vertex (w1)   at (-1, -1.3);
      \vertex (w2)   at ( 1, -1.3);
      \vertex (out2) at ( 2, -1.3);
      \vertex [dot] (b) at (-1,0);
      \vertex [dot] (d) at ( 1,0);
      \diagram*{
        (v1) -- [very thick, momentum'=\(p_1\)] (in1),
        (v1)-- [very thick, momentum=\(k_0\)] (v2) -- [very thick, momentum=\(p_4\)] (out1),
        (w1) -- [very thick, momentum=\(p_2\)] (in2),
        (w1)-- [very thick, momentum'=\(k_2\)] (w2) -- [very thick, momentum'=\(p_3\)] (out2),
        (v1) -- [photon, double, momentum'=\(q_1\)] (b) -- [photon, double, momentum'=\(q_2\)] (w1),
        (v2) -- [photon, double, momentum=\(q_3\)] (d) -- [photon, double, momentum=\(q_4\)] (w2),
        (b)  -- [photon, draw=blue, momentum=\(k_1\)] (d)
      };
      \node[purple] at ($ (v1)!0.5!(v2) $) {\textbar};
      \node[purple] at ($ (w1)!0.5!(w2) $) {\textbar};
      \node[purple] at ($ (b)!0.5!(d) $)   {\textbar};
    \end{feynman}
  \end{tikzpicture}
  \hspace{1.5em} 
  $\Longrightarrow$
  \hspace{1.5em}
  \begin{tikzpicture}[scale=0.3,baseline=(intC.center)]
    \begin{feynman}
      \vertex[] (int0);
      \vertex[above=.5cm of int0]  (int-1);
      \vertex[below=1.4cm of int0] (intC);
      \vertex[below=1.4cm of intC] (int1);
      \vertex[below=.5cm of int1]  (int2);
      \vertex[right=1cm of intC] (up);
      \vertex[left=1cm  of intC] (down);
      \diagram*{
        (int-1) -- [thick, momentum'={\(q\)}] (int0),
        (int1)  -- [thick, momentum'={\(q\)}] (int2),
        (int0)  -- [photon, momentum=\(q_3\)] (up)   -- [photon, momentum=\(q_4\)] (int1),
        (int0)  -- [photon, momentum'=\(q_1\)] (down) -- [photon, momentum'=\(q_2\)] (int1),
        (down)  -- [photon, draw=blue, momentum=\(k_1\)] (up)
      };
      \node[purple] at ($ (int0)!0.5!(int1) $) {\textbar};
    \end{feynman}
  \end{tikzpicture}
\caption{Vertical and horizontal wavy lines denote Glauber and soft gravitons, respectively.
  Integrating over rapidities contracts the massive lines to effective vertices through which
  the momentum $q$ flows.
  }
\label{fig:Hdiagram}
\end{figure}
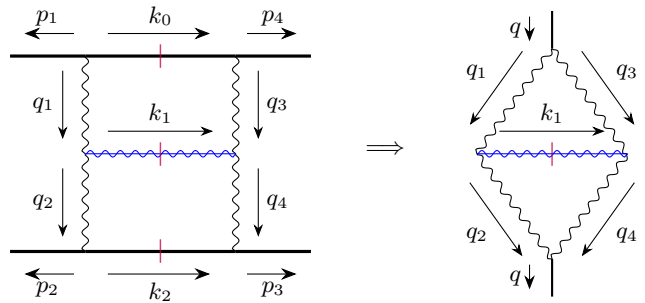

The $s$-channel discontinuity of the two-loop $2\rightarrow 2$ amplitude is proportional to the imaginary part of the amplitude, which is determined by the three-particle cut involving the square of two tree-level $2\rightarrow 3$ amplitudes,
\begin{align}
\label{eq:3pc}
&2 \mathrm{Im}\,\mathcal{M}^{(2)}_{2\rightarrow 2}(s,q^2)=\int \mathrm{d}\mathcal{P}_{3} \,|\mathcal{M}_{2\rightarrow 3}^{(0)}|^2,\\
&\quad \mathrm{d}\mathcal{P}_{3} = \mathrm{d}\Phi(k_0,k_1,k_2) \,\hat{\delta}^{(D)}\big(p_1+p_2+\sum_{i=0}^2k_i\big) \,,  \nonumber 
\end{align}
with the phase space measure defined as $\mathrm{d}\Phi(k_0,k_1,k_2)=\prod_{i=0}^2 \hat{\mathrm{d}}^D k_i \,\delta^+(k_i^2)$ and where the on-shell momenta satisfy $p_1^2=k_0^2=p_4^2=0$, $p_2^2=k_2^2=p_3^2=0$ and $k^2_1=0$. Using the conventions in Fig.\ref{fig:Hdiagram}, the overall momentum conservation implies $p_4=-p_1-q$ and $p_3=-p_2+q$ and thus $q_3=q-q_1$ and $q_4=q-q_2$, with $t=q^2\simeq-q^2_{\perp}$.

Combining MRK approximations for the tree-level $2\rightarrow 3$ amplitudes and for the three-particle phase space~\cite{DelDuca:1995hf},
\begin{align}
\int \mathrm{d}\mathcal{P}_{3}\simeq\frac{1}{4s}\bigg(\frac{\log(s)}{2\pi}\bigg)
\int_{q_{1 \perp}, q_{2 \perp}}\,,
\end{align} 
one can integrate out the rapidities obtaining
\begin{subequations}\label{eq:H1-def}
\begin{align}
&2 \mathrm{Im}\,\mathcal{M}^{(2)}_{2\rightarrow 2}(s,q^2)\simeq\frac{(8\pi G_N)^3 s^3}{8\pi}\log (s)H_1({q}_{\perp}^2),\\&
\hspace{-6pt}H_1(q^2_{\perp})\equiv \varsigma^{2\epsilon} \int_{q_{1 \perp},\,q_{2 \perp}}\frac{\mathcal{K}^{\mathrm{GR}}({q}_{1},{q}_{2};{q})}{q_{1 \perp}^2q_{2 \perp}^2(q-q_1)_{\perp}^2(q-q_2)_{\perp}^2}.
\end{align}
\end{subequations}
The term $\varsigma=\mu^{2}\exp(\gamma_E)$ fixes the renormalization scheme.
Notice that the appearance of $\log(s)$ is entirely due to the integration over the soft graviton's rapidity and that in MRK the computation of $2 \mathrm{Im}\,\mathcal{M}^{(2)}_{2\rightarrow 2}(s,q^2)$ drastically simplifies from a two-loop four-point integral in $D$ dimensions to a two-loop massless two-point integral in $d=D-2$ dimensions, graphically represented in Fig. \ref{fig:Hdiagram}, which is significantly easier. By power counting, the result must be proportional to $(q_{\perp}^2)^{d-2}$.

Using IBPs~\cite{Lee:2012cn}, one can reduce $H_1(q^2_{\perp})$ to two simple master integrals which are product and iterations of bubbles in \eqref{eq:generalised-bubbles}. The $\epsilon$ expansion of the discontinuity reads in momentum and impact-parameter space, respectively:
\begin{subequations}
\begin{align}
    &2 \mathrm{Im}\,\mathcal{M}^{(2)}_{2\rightarrow 2}(s,q^2)= 8 G_N^3 s^3\log(s) \left(\frac{4\pi\mu^2}{q^2}\right)^{2\epsilon} \\
    &\qquad \qquad \qquad \qquad \quad \times \Big(-\frac{1}{\epsilon^2}+\frac{2}{\epsilon}+\zeta_2+\mathcal{O}(\epsilon^0)\Big), \nonumber \\
    &2 \mathrm{Im}\,\tilde{\mathcal{M}}^{(2)}_{2\rightarrow 2}(s,b^2)=\frac{8 G_N^3 s^2}{\pi b^2}\log(s) \left(  \pi b^2 \varsigma e^{\gamma_E} \right)^{3\epsilon} \\
    & \qquad \qquad \qquad \qquad \quad \times \Big(-\frac{1}{\epsilon}+2+\mathcal{O}(\epsilon^0)\Big). \nonumber
\end{align}
\end{subequations}
Notice that this identically agrees with eq.(3.7) of~\cite{DiVecchia:2020ymx}. 

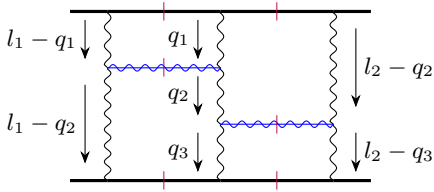
\begin{figure}[t]
  \centering
  \begin{tikzpicture}
    \centering
    \begin{feynman}
        \vertex[] (up0);
        \vertex[left=0.5cm of up0] (up-1);
        \vertex[right=1.5cm of up0] (up1);
        \vertex[right=1.5cm of up1] (up2);
        \vertex[right=0.5cm of up2] (up3);

        \vertex[below=2.25cm of up0] (low0);
        \vertex[left=0.5cm of low0] (low-1);
        \vertex[right=1.5cm of low0] (low1);
        \vertex[right=1.5cm of low1] (low2);
        \vertex[right=0.5cm of low2] (low3);

        \vertex[below=.75cm of up0] (int0);
        \vertex[below=.75cm of up1] (int1up);
        \vertex[below=.75cm of int1up] (int1low);
        \vertex[below=1.5cm of up2] (int2);
        \diagram*{
          (up-1) -- [very thick] (up0) -- [very thick] (up1) -- [very thick] (up2) -- [very thick] (up3),
          (low-1) -- [very thick] (low0) -- [very thick] (low1) -- [very thick] (low2) -- [very thick] (low3),
          (up0) -- [photon, double, momentum'=\(l_1-q_1\)] (int0) -- [photon, double, momentum'=\(l_1-q_2\)] (low0),
          (up1) -- [photon, double, momentum'=\(q_1\)] (int1up) -- [photon, double, momentum'=\(q_2\)] (int1low) -- [photon, double, momentum'=\(q_3\)] (low1),
          (up2) -- [photon, double, momentum=\(l_2-q_2\)] (int2) -- [photon, double, momentum=\(l_2-q_3\)] (low2),
          (int0) -- [photon, draw=blue] (int1up),
          (int1low) -- [photon, draw=blue] (int2)
        };
        \node[purple] at ($ (up0)!0.5!(up1) $) {\scriptsize\textbar};
        \node[purple] at ($ (up1)!0.5!(up2) $) {\scriptsize\textbar};
        \node[purple] at ($ (low0)!0.5!(low1) $) {\scriptsize\textbar};
        \node[purple] at ($ (low1)!0.5!(low2) $) {\scriptsize\textbar};
        \node[purple] at ($ (int0)!0.5!(int1up) $) {\scriptsize\textbar};
        \node[purple] at ($ (int1low)!0.5!(int2) $) {\scriptsize\textbar};
    \end{feynman}
  \end{tikzpicture}
  
  \caption{Momenta parametrisation for $H_2$.}
\label{fig:h2sixdiagrams}
\vspace{-10pt}
\end{figure}

\paragraph*{Double-H diagrams:} In general, at four-loops there are six diagrams contributing to the 5PM amplitude at leading logarithmic accuracy, in agreement with the counting \eqref{eq:counting_top}. However, it turns out they are all equal and equivalent to the one in Fig.~\ref{fig:h2sixdiagrams}, where two horizontal soft gravitons are simultaneously cut in the Regge limit.
Such equivalence is a direct consequence of target-projectile symmetry and factorization properties of the high-energy scattering.
Using again the MRK approximation and accounting for both the eikonal symmetry factor from the Glauber graviton exchanges and the rapidity–ordering factor from the two soft emissions,  one finds
\allowdisplaybreaks
\begin{subequations}
\begin{align}
&2 \mathrm{Re}\,\mathcal{M}_{2\rightarrow2}^{(4)}(s,q^2)\simeq -\frac{(8\pi G_N)^{5}s^4}
{64\pi^2}\log^2(s)H_2(q^2_{\perp}),\\\nonumber
\label{eq:H2}
&H_2(q^2_{\perp})\equiv \varsigma^{4\epsilon} \int_{l_{1 \perp},\,l_{2 \perp},\,q_{2 \perp}} J(q_2,l_1)\,q_2^2\,J(q_2,l_2) \\&\hspace{2.5cm}\times \hat{\delta}^{(d)}(l_{1 \perp}+l_{2 \perp}+q_{\perp}-q_{2 \perp}),
\end{align}
\end{subequations}
where we used the momentum parametrisation in Fig. \ref{fig:h2sixdiagrams} and we have defined
\begin{align}
\label{eq:J}
\hspace{-10pt}J(q,l)&=\int_{k_{\perp}}\frac{\mathcal{K}^{\mathrm{GR}}(k,q;l)}{k_{\perp}^2(k-l)^2_{\perp}q_{\perp}^2(q-l)^2_{\perp}} =\sum_{i=1,2} J_i(q,l),
\end{align}
with $J_i(q,l)$ given in App.~\ref{sec:app1}. Therefore, one can decompose $H_2(q^2_{\perp}) = H_2^{(A)}(q_{\perp}^2)+H_2^{(B)}(q_{\perp}^2)$ where we define
\begin{align}
&H_2^{(A)}(q^2_{\perp})=\varsigma^{4\epsilon}\int_{l_{1 \perp},l_{2 \perp},q_{2 \perp}} \hat{\delta}^{(d)}(l_{1 \perp}+l_{2 \perp}+q_{\perp}-q_{2 \perp}) \nonumber\\
&\hspace{0.4cm}\times q_2^2 \,(J_2(q_2,l_1)J_2(q_2,l_2)+2J_1(q_2,l_1)J_2(q_2,l_2)), \label{eq:h2a-def}\\
&H_2^{(B)}(q^2_{\perp})=\varsigma^{4\epsilon}\int_{l_{1 \perp },l_{2 \perp},q_{2 \perp}}\hat{\delta}^{(d)}(l_{1 \perp}+l_{2 \perp}+q_{\perp}-q_{2 \perp}) \nonumber\\
&\hspace{1.6cm}\times q_2^2 \, J_1(q_2,l_1)J_1(q_2,l_2). \label{eq:h2b-def}
\end{align}
The first integral $H_2^{(A)}(q^2_{\perp})$ can be solved in terms of bubbles \eqref{eq:generalised-bubbles} and tensor bubbles \eqref{eq:tensor-bubbles} integrals, see App.~\ref{sec:app1} for further details. Interestingly, the second integral $H_2^{(B)}(q^2_{\perp})$ is proportional to 
 \begin{align}\label{eq:kite-h2}
\hspace{-10pt}H_2^{(B)}(q^2_{\perp}) \propto \int_{l_{1 \perp}, l_{2 \perp}} \!\!\frac{(l_{1 \perp}^2)^\frac{d}{2}(l_{2 \perp}^2)^\frac{d}{2}}{(q-l_1)^2_{\perp}(q-l_2)^2_{\perp}(q-l_1-l_2)^2_{\perp}},  
 \end{align}
which is a scalar kite topology with non-integer exponents, similar to the one found at four-loops in QCD~\cite{Caron-Huot:2013fea}. As shown in App.~\ref{sec:app1}, it can be solved~\cite{Kotikov:2013eha} in $d$ dimensions using the Gegenbauer polynomial technique~\cite{Chetyrkin:1980pr,Kotikov:1995cw,Kotikov:2018wxe} in terms of the hypergeometric function ${}_3F_2$.

Summing together the six diagrams contributing to the 5PM order we get the following $\epsilon$-expanded results:
\begin{align}
     & 2 \mathrm{Re} {\mathcal{M}}_{2\to2}^{(4)}(s,q^2) \simeq - 4\,G_N^5 s^4 \log^2(s) \frac{q^2}{\pi} \left(\frac{4 \pi \mu^2}{q^2}\right)^{4\epsilon} \nonumber \\
     &\qquad\quad \times\left[-\frac{1}{\epsilon^3}-\frac{1}{\epsilon^2}-\frac{9}{\epsilon}+\frac{2}{\epsilon} \zeta_2+\frac{2}{\epsilon} \zeta_3+O(1)\right]\,, \label{eq:h2-result}\\
     & 2 \mathrm{Re}\tilde{\mathcal{M}}_{2\to2}^{(4)}(s,b^2) \simeq - 256\,G_N^5 \frac{ s^3 \log^2 (s)}{\left(\pi b^2\right)^2} \left(\pi b^2 \varsigma e^{\gamma_E} \right)^{5\epsilon} \nonumber \\
     &\qquad\quad \times \left[\frac{1}{8 \epsilon^2}-\frac{1}{\epsilon}+\frac{5}{2}+\frac{5}{16} \zeta_2-\frac{1}{4} \zeta_3+O(\epsilon)\right]\,. \label{eq:h2-result-bspace}
\end{align}

\section{Predicting sub-leading logs from leading logs through analyticity and crossing}
\label{sec:disp_rel}

The computation of the leading classical double logarithm at 5PM allows us to extract further information on the amplitude at the same loop order by using $s\leftrightarrow u$ crossing–symmetric dispersion relations. In particular, the result we obtained is related to the single classical logarithm in the imaginary part of $\mathcal{M}_{2\to2}^{(4)}(s,t)$, which will be our starting point to get information on the eikonal phase. Crossing-symmetric constraints of this type were first identified at 3PM in the Regge limit~\cite{Amati:1990xe} and later extended to massive scattering in~\cite{DiVecchia:2020ymx}. It was further shown in QCD~\cite{Caron-Huot:2017fxr} that crossing symmetry dictates how imaginary terms appear in the leading expansion in $t/s$. In particular, when logarithms are written in the combination $\log|s/t| - i\pi/2$, their coefficients are real (imaginary) for amplitudes with minus (plus) crossing signature. Here we generalize the result of~\cite{Caron-Huot:2017fxr} to gravity, or equivalently extend the analysis of~\cite{Amati:1990xe} to higher PM orders. 

Since we consider scalar scattering, the amplitude has positive crossing signature. As discussed above, the scaling of the gravitational coupling as $G s$ implies that we must retain all orders in $t/s$. For this reason, we replace $s/t$ by the crossing-covariant variable $z_t = 1 + 2s/t$, whose powers encode the correct transformation properties under $s\leftrightarrow u$ crossing. The same variable enters the natural crossing-symmetric definition of the logarithm, $L \equiv \left(\log(-z_t)+\log(z_t)\right)/2= \log|z_t| - i\pi/2$. In App.~\ref{app:massless-appendix} we use dispersion relations and crossing symmetry to derive the following form of the amplitude,
\begin{equation}\label{eq:ampl-high-energies}
  \mathcal{M}_{2\rightarrow 2}^{(\ell)} = \frac{G_N^{\ell+1}(tz_t)^{\ell + 2}}{t} \sum_{j=0}^\infty\sum_{k = 0}^{\text{min}(\ell,j)}i^{\ell + j}f^{(\ell)}_{(j,k)}(t) z_t^{-j}\,L^k \,,
\end{equation}
where the real coefficients $f^{(\ell)}_{(j,k)}$ scale as $(\mu^{2}/t)^{\ell\epsilon}$.  
In this representation all $s$-dependence is explicit, appearing only through $L$ or the overall powers of $z_t$. Combining this structure with the assumption of eikonalization,\footnote{As noted, for the massless case of interest the known violations of strict eikonalization do not affect our result.} we obtain a prediction for the single radiative logarithm.

To do so, it is useful to expand $\delta_{\text{Cl}}$ and $\Delta_Q$ in powers of $\log(s/|t|)$. The same counting rules for the amplitude apply to these at each order in $G_N$.  In particular, for the 3PM and 5PM classical phase, we write
\begin{equation}
    \delta^{(2j )}_{\text{Cl}} = \sum_{k=0}^{j}\delta^{(2j), k}_{\text{Cl}}(b)\log^k\left({s}/{|t|}\right),
\end{equation}
with a similar expansion for $\Delta_Q$. Equating Eqs. (\ref{eq:ampl-high-energies}) and (\ref{Amp Eik}), we may solve for the unknown coefficients $f^{(\ell)}_{(j,k)}$ in terms of $\delta_{\text{Cl}}$ and $\Delta_Q$.  Doing so then gives the following relations between the real and imaginary parts of the classical eikonal phase at two loops,
\begin{align}
    \text{Re}\,\delta_{\text{Cl}}^{(2),0} &= \frac{\pi}{2}\text{Im}\,\delta_{\text{Cl}}^{(2),1} -8\epsilon^2 (\delta^{(0)}_{\text{Cl}})^3,\label{3PM}
\end{align}
and at four loops,
\begin{align}
    \text{Im}\,\delta^{(4),1}_{\text{Cl}} &= 16\epsilon(1-4\epsilon)(\delta^{(0)}_{\text{Cl}})^2 \text{Im}\,\delta^{(2),1}_{\text{Cl}}-\pi\text{Re}\,\delta^{(4),2}_{\text{Cl}}. \label{5PM}
\end{align}
Such a relation at 3PM has been previously derived in~\cite{Amati:1990xe, DiVecchia:2020ymx}, and it is identical to Eq. (\ref{3PM}) if one uses the fact that $ \delta^{(2j )}_{\text{Cl}} \sim (b^2)^{(2j  +1)\epsilon }$ in dimensional regularization.  It should be mentioned that the formulae given above are only valid in dimensional regularization; however, expressions valid for other IR regulators may be obtained from the procedure described above.  We also note that in this work we have only obtained 
\begin{equation}
    \text{Re}\,\delta^{(4),2}_{\text{Cl}} = \frac{1}{2} \mathrm{Re}\,\tilde{\mathcal{M}}_{2\to2}^{(4)}(s,b^2) \,,
\end{equation}
from the above calculations at 5PM using \eqref{Amp Eik}. We then obtain $\text{Im}\,\delta^{(4),1}_{\text{Cl}} $ from the above relations \eqref{5PM}. Using the known results for $\delta^{(2)}_{\text{Cl}}$ to $O(\epsilon)$ in Refs.~\cite{DiVecchia:2019kta, Rothstein:2024nlq}, we find
\begin{equation}
\label{eq:Imdelta4}
\hspace{-10pt}\text{Im}\,\delta^{(4),1}_{\text{Cl}} = \frac{16}{\pi}\left(\pi b^2 \varsigma e^{\gamma_E} \right)^{5\epsilon}\left[-\frac{1}{\epsilon}-\zeta_3 + 6 + O(\epsilon)\right]\,.
\end{equation}
It is curious that, similar to the case at 3PM, there is a cancellation between the leading $1/\epsilon^2$ divergences in the dispersion relation between the two terms. Although we derived Eq. \eqref{eq:Imdelta4} using only analyticity and crossing symmetry, it would be interesting to rederive it independently through an MRK discontinuity cut or the RRGE method described above. We leave this question for future investigations.

It is important to note that this result along with \eqref{eq:h2-result-bspace} must agree with the high energy limit of  the massive case. In the EFT this is expected: the leading logarithms are fixed entirely by the soft sector, which is insensitive to the particle masses, while subleading contributions from collinear radiation do depend on the mass.

Finally, the EFT also implies additional consistency relations for rapidity anomalous dimensions~\cite{Rothstein:2023dgb}. In QCD, an infinite tower of relations between soft functions with different indices was derived in~\cite{Rothstein:2024fpx}, and analogous structures are expected to hold in gravity.

\section{Conclusions and future directions}\label{sec:conclusion}

The behavior of gravitational amplitudes in the Regge limit encodes rich physics, both theoretically -- as an effective field theory description -- and phenomenologically, as it underlies resummations relevant to the two–body problem in the ultra-relativistic regime. Yet, compared to gauge theory where Regge methods have long been a powerful tool for strong interactions, the gravitational case remains largely unexplored at higher-loop order.  

Building on Lipatov's seminal work, the eikonal formalism, and recent advances in soft–collinear effective theory, we have initiated a systematic study of classical leading–logarithmic contributions to the $2\!\to\!2$ amplitude of massive particles in gravity. The resulting series of multi–H diagrams is the main subject of this work.  

We begin by revisiting the imaginary part of the single H diagram at 3PM order, confirming previous results, and we then compute for the first time the leading–logarithmic contribution of the real part of the double–H diagram at 5PM order. This requires the evaluation of a non-trivial topology involving the massless two–point function at higher-loop order, which also appears in the calculation of four–loop anomalous dimensions in QCD~\cite{Caron-Huot:2013fea}. For all diagrams, the SCET framework and the multi–Regge expansion produce the same integrand representation, showing complete agreement.

Finally, we develop a new set of dispersion relations for the gravitational scattering amplitude in the Regge limit, combining signature symmetry in the ultra–relativistic regime~\cite{Caron-Huot:2017fxr} with eikonal exponentiation~\cite{Amati:1990xe,DiVecchia:2020ymx}, to extract some pieces of the ultra-relativistic limit of the classical eikonal phase. In this way we recover both the real and imaginary parts of the 3PM eikonal phase in the ultra-relativistic limit found in~\cite{Amati:1990xe,DiVecchia:2019kta}, and determine the leading 5PM real contribution from the double–H diagrams together with the subleading logarithmic correction to the imaginary part. Despite the fact that we work in the massless limit, our results are valid in the high energy limit of the massive case at 5PM order.

Our work opens several avenues for future investigation. An immediate direction is the study of the triple–H diagram and higher–loop contributions, with the long–term goal of achieving an all–order resummation of the $2\!\to\!2$ amplitude in the Regge limit. Moreover, our results provide useful input to the gravitational S-matrix bootstrap program~\cite{Haring:2022cyf,Haring:2024wyz}, as the Regge–cut contributions directly inform the non-perturbative analytic structure of the elastic amplitude. We also plan to study the contribution of these effects to the corresponding classical observables, such as the scattering angle, and to extend the methods presented here beyond the leading–logarithmic approximation, paving the way for a systematic resummation in the ultra-relativistic regime.

Finally, the high-energy limit explored here may also find phenomenological applications. In particular, it could inform self–force and EOB–based resummations of scattering observables~\cite{Damour:2022ybd,Buonanno:2024byg}, where logarithmic growth at high energy~\cite{Dlapa:2022lmu,Bini:2022enm} challenges current resummation schemes for the scattering angle when compared to numerical simulations~\cite{Swain:2024ngs,Long:2025nmj}. Our EFT approach suggests a physically motivated resummation strategy that systematically organizes these logarithms, potentially shedding light on the high–energy puzzle~\cite{DEath:1976bbo,Kovacs:1977uw,Kovacs:1978eu,Gruzinov:2014moa,Ciafaloni:2015xsr,Ciafaloni:2018uwe,DiVecchia:2022nna,Gonzo:2023cnv}. We hope to come back to this problem in the near future.

\textit{Acknowledgments---} We thank J.~Parra-Martinez, M.~Zeng and S.~Zhiboedov for useful discussions. 
The work of R.G.\ is supported by the Royal Society grant RF\textbackslash ERE\textbackslash 231084.
I.Z.R. is supported by the Department of Energy (Grant No. DE- FG02-04ER41338 and FG02-06ER41449).
M.S. is supported in part by the U.S. Department of Energy (DOE) under award number DE SC0009937, and by the European Research Council (ERC) Horizon Synergy Grant “Making Sense of the Unexpected in the Gravitational Wave Sky” grant agreement no. GWSky–101167314.  We are also grateful to the Mani L. Bhaumik Institute for Theoretical Physics for support.

\appendix

\section{Details of the master integral evaluation and dispersion relations}
\label{sec:app1}

\subsection{Relevant master integrals}
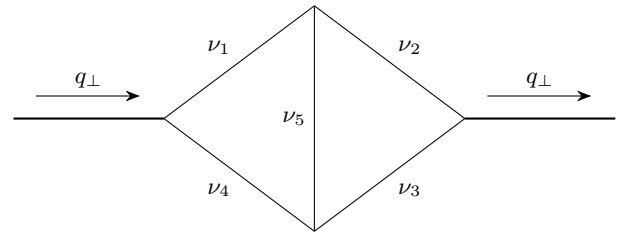
\begin{figure}[h!]
\centering
  \begin{tikzpicture}[scale=0.1,baseline=(a.center)]
    \begin{feynman}
        \vertex[] (a);
        \vertex[left=2cm of a] (in);
        \vertex[right=2cm of a] (cntr);
        \vertex[right=2cm of cntr] (b);
        \vertex[right=2cm of b] (out);
        \vertex[above=1.5cm of cntr] (up);
        \vertex[below=1.5cm of cntr] (low);
        \diagram*{
          (in) -- [thick,momentum=\(q_{\perp}\)] (a) -- [edge label=\(\nu_1\)] (up) -- [edge label=\(\nu_2\)] (b) -- [thick,momentum=\(q_{\perp}\)] (out),
          (a) -- [edge label'=\(\nu_4\)] (low) -- [edge label'=\(\nu_3\)] (b),
          (up) -- [edge label'=\(\nu_5\)] (low)
        };
    \end{feynman}
  \end{tikzpicture}
  \caption{The scalar massless kite integral topology $K$.}
  \label{fig:kite}
\end{figure}

The {\it generalised bubble  } integrals are defined as
\begin{subequations}
\begin{align}
\label{eq:generalised-bubbles}
    &\Lambda_{a,b}(q_{\perp}) = \int_{l_{\perp}} \frac{1}{[(q-l)_{\perp}^2]^a[l^2_{\perp}]^b} = c_{a,b} (q^2_{\perp})^{\frac{d}{2}-a-b}, \nonumber\\
    &c_{a,b} = \frac{1}{(4\pi)^{\frac{d}{2}}}\frac{\Gamma(a+b-\frac{d}{2})}{\Gamma(a)\Gamma(b)}\frac{\Gamma(\frac{d}{2}-a)\Gamma(\frac{d}{2}-b)}{\Gamma(d-a-b)},
\end{align}
\end{subequations}
whereas the {\it vector bubble} integrals are
\begin{align}\label{eq:tensor-bubbles}
    \nonumber\Lambda^i_{a,b}(q_{\perp}) = &\int_{l_{\perp}} \,\frac{l^i_{\perp}}{[(q-l)^2_{\perp}]^a[l^2_{\perp}]^b} = \frac{q^i_{\perp}}{2}[\Lambda_{a,b}(q_{\perp})\\&+\frac{1}{q^2_{\perp}}\big( \Lambda_{a,b-1}(q_{\perp}) - \Lambda_{a-1,b}(q_{\perp}) \big)].
\end{align}
Using eq. \eqref{eq:generalised-bubbles} one gets for $J(q,l)$ in \eqref{eq:J}
\allowdisplaybreaks
\begin{align}
\label{eq:Jdecomposition}
&J(q,l)=J_1(q,l)+J_2(q,l),
\\&J_1(q,l) = c_{1,1} \frac{(l^2_{\perp})^\frac{d}{2}}{q^2_{\perp}(q-l)^2_{\perp}},\\\nonumber
&J_2(q,l)=\left[\frac{d}{2} - \frac{d}{2}\frac{l^2_{\perp}}{q^2_{\perp}}-\left(1-\frac{d}{2}\right)\frac{(l-q)^2_{\perp}}{q^2_{\perp}}\right]\Lambda_{1,1}(l_{\perp}-q_{\perp})  \\  &\hspace{2cm}+(q_{\perp}\leftrightarrow l_{\perp}-q_{\perp}).
\end{align}
The result for $H_2^{(A)}(q^2_{\perp})$ in equation \eqref{eq:h2a-def} can be evaluated using the bubble integrals above, getting
\begin{align}
    H_2^{(A)}(q) &= \frac{\varsigma^{4\epsilon}}{(q_{\perp}^2)^3}\left(\frac{q_{\perp}^2}{4\pi}\right)^{2d} g^{(d)}  \Big[f_1^{(d)}+f_2^{(d)}\big(h_a^{(d)}+h_b^{(d)}\big)\Big]\,, \nonumber 
\end{align}
with the coefficients 
\begin{align}
    g^{(d)} & = \frac{\pi  2^{5-4 d} \Gamma \left(\frac{d}{2}\right)^2}{2\Gamma \left(\frac{d-1}{2}\right)^2 \Gamma \left(\frac{5 d}{2}-3\right)}\,, \\
    f_1^{(d)} & = 4^{d+1} \left(d \left(5 d^2+d-9\right)-6\right) \nonumber \\
    & \qquad \qquad \times \Gamma (3-2 d) \Gamma (d-2)^2 \Gamma \left(\frac{d}{2}\right)\,, \\
    f_2^{(d)} & = \Gamma \left(1-\frac{d}{2}\right) \Gamma \left(\frac{d}{2}-1\right)\,, \\
    h_a^{(d)} & = -2^{2 d+1} (d (2 d-1)-4) \\
    & \times  \frac{\Gamma (1-2 d) \Gamma (1-d) \Gamma \left(\frac{d}{2}-1\right) \Gamma (d) \Gamma (2 d)}{\Gamma \left(1-\frac{3 d}{2}\right)\Gamma\left(\frac{3 d}{2}\right)}\,, \nonumber \\
    h_b^{(d)} & = -\frac{1}{\sqrt{\pi}}3 ((d-3) d+4) (d (3 d-2)-4) \\
    & \times  \Gamma \left(\frac{3}{2}-d\right) \Gamma \left(1-\frac{d}{2}\right) \Gamma \left(\frac{d}{2}\right) \Gamma \left(\frac{3 d}{2}-3\right)\,, \nonumber
\end{align}
Both $H_2^{(A)}(q)$ and $H_2^{(B)}(q)$ exhibit a leading $1/\epsilon^4$ behavior that cancels out in their sum in eq.\eqref{eq:h2-result}. We express the integral in \eqref{eq:kite-h2} as $H_2^{(B)}(q_{\perp}^2)=\varsigma^{4\epsilon}(c_{1,1})^2K\left(-d/2,1,-d/2,1,1\right)$, making use of a generalized massless kite topology $K$ defined as:
\begin{align}\label{eq:general-kite}
    &\hspace{-10pt}K(\nu_1,\nu_2,\nu_3,\nu_4,\nu_5) \\
    &= \int_{l_{1 \perp},l_{2 \perp}} \frac{(l_{1 \perp})^{-2\nu_1}(l_{2 \perp})^{-2\nu_3}}{[(l_2-q)_{\perp}]^{2\nu_2}[(l_1-q)_{\perp}]^{2\nu_4}[(l_1+l_2-q)_{\perp}]^{2\nu_5}}  \nonumber 
\end{align}
and schematically represented in Fig.\ref{fig:kite}.
The integral $K$ has been solved in~\cite{Kotikov:2013eha} using the Gegenbauer polynomial technique~\cite{Kotikov:1995cw,Kotikov:2018wxe} and in~\cite{Bierenbaum:2003ud} through a double Mellin Barnes representation. In~\cite{Bierenbaum:2003ud} the authors proved that the expansion for $\epsilon\to 0$ of such integral involves only rational numbers and multiple Zeta values. Using result in appendix B of~\cite{Kotikov:2016rgs}, we obtain 
\begin{equation}
  \hspace{-10pt}  H_2^{(B)}(q) = \frac{\varsigma^{4\epsilon}}{(q_{\perp}^2)^3}\left(\frac{q_{\perp}^2}{4\pi}\right)^{2d} \Big[\bar f_1^{(d)}+\bar f_2^{(d)}\big(\bar h_a^{(d)}+\bar h_b^{(d)}\big)\Big],
\end{equation}
with the following coefficients 
\begin{widetext}
\vspace{-5pt}
\begin{align}
    & \bar f_1^{(d)} = \frac{8 \Gamma \left(1-\frac{3 d}{2}\right) \Gamma (1-d) \Gamma \left(1-\frac{d}{2}\right) \Gamma \left(\frac{d}{2}\right)^4 \Gamma (d) \Gamma \left(\frac{3 d}{2}\right)}{\Gamma
   (d-1) \Gamma \left(\frac{5 d}{2}-3\right)}\,, \qquad \qquad \bar f_2^{(d)} = \frac{(d-2) \Gamma \left(1-\frac{d}{2}\right)^5 \Gamma \left(\frac{d}{2}\right)^5}{\Gamma \left(2-\frac{d}{2}\right)^4 \Gamma \left(-\frac{d}{2}\right)}\,, \nonumber\\
    & \bar h_a^{(d)} = -\frac{(d-2)^2 \Gamma (1-2 d) \Gamma (2 d) \, _3F_2\left(1,d-2,\frac{3 d}{2}-2;1-\frac{d}{2},\frac{3 d}{2}-1;1\right)}{(3 d-4) \Gamma (2 d-2) \Gamma \left(\frac{5 d}{2}-3\right)}\,, \nonumber\\
    & \bar h_b^{(d)} = -\frac{\Gamma (1-d)^2 \Gamma \left(1-\frac{d}{2}\right) \Gamma \left(\frac{d}{2}\right) \Gamma (d)^2 \, _3\tilde{F}_2\left(d-2,\frac{3 d}{2}-2,\frac{5 d}{2}-3;\frac{3 d}{2}-1,2
   d-3;1\right)}{\Gamma \left(1-\frac{3 d}{2}\right) \Gamma (d-2) \Gamma \left(\frac{3 d}{2}\right)}\,.
\end{align}
\vspace{-5pt}
\end{widetext}

\section{Dispersion relation constraints for gravity amplitudes in the Regge limit}\label{app:massless-appendix}

The fixed-$t$ dispersion relation in Mellin space for a crossing-symmetric amplitude is given by
\begin{align}\label{disp}
&\mathcal{M}_{2\rightarrow 2}(s,t) = \frac{1}{\pi}\int_{\gamma-i\infty}^{\gamma+i\infty} d\omega\,
\frac{a(t,\omega)}{\sin(\pi\omega)}\\ 
& \qquad \qquad \times \Big[ \!\left(\frac{-s-i\epsilon}{-t}\right)^{\omega} +\!\left(\frac{s+t-i\epsilon}{-t}\right)^{\omega} \Big] \nonumber
\end{align}
where we have defined the $\epsilon'$-regularized Mellin transform of the $s$-channel discontinuity
\begin{equation}
\label{eq:disc_a}
a(t,\omega)= \int_{-t}^\infty \frac{ds^\prime}{s^\prime} \text{Disc}_s\mathcal{M}_{2\rightarrow 2}(s^\prime,t) \left(\frac{s^\prime}{-t}\right)^\omega e^{-\epsilon' s^\prime},
\end{equation}
and we have taken the subtraction point to be $-t$. The poles in $\epsilon'$ correspond to local (real) counterterms, which play no role in our analysis since we make no claims regarding contact interactions.

Now we change the variables from $(s,t)$ to $(z_t,t)$ where 
\begin{equation}
z_t= \frac{2s}{t}+1
\end{equation}
is crossing odd, and re-write the dispersion relation in terms of the crossing even $L \equiv  \log |z_t| - i\pi/2$
\begin{align}
&\mathcal{M}_{2\rightarrow 2}(z_t,t)\!=\! \frac{1}{\pi}\int_{\gamma-i\infty}^{\gamma+i\infty} \!\!d\omega\,\frac{a(t,\omega)}{\sin(\pi\omega)} e^{\omega L} \\
&\qquad\times \bigg\{
\cos\left(\frac{\pi\omega}{2}\right)\left[\left(\tfrac{1+1/z_t}{2}\right)^{\omega}\!+\! \left(\tfrac{1-1/z_t}{2}\right)^{\omega}\right]\nonumber + \\
&\qquad \hspace{5pt} +i \sin\left(\frac{\pi\omega}{2}\right)\left[\left(\tfrac{1-1/z_t}{2}\right)^{\omega}-\left(\tfrac{1+1/z_t}{2}\right)^{\omega}\right]\bigg\}\nonumber,
\end{align}
where $a(t,\omega)$ is now written as a polynomial in $z_t$. We can then re-write the dispersion relation as \\
\begin{align}
\label{final}
\mathcal{M}_{2\rightarrow 2}(z_t,t)&= \frac{1}{\pi}\int_{\gamma-i\infty}^{\gamma+i\infty} \frac{d\omega}{2^\omega}\,
\frac{a(t,\omega)}{\sin(\pi\omega)} e^{ \omega L} \\
& \times \bigg[\cos\left(\frac{\pi\omega}{2}\right)\sum_{n}\binom{\omega}{2n} |z_t|^{-2n}\nonumber \\
&\qquad +i\sin\left(\frac{\pi\omega}{2}\right)\sum_{n}\binom{\omega}{2n+1} |z_t|^{-2n-1}\bigg]\,. \nonumber
\end{align}
This manifestly crossing-invariant representation ties the powers of $i$ in each term to the corresponding power of $z_t$. To obtain our expression~\eqref{eq:ampl-high-energies}, we must establish both the reality properties of the coefficients $f_{(j,k)}^{(\ell)}$ and the structure of the powers of $L$. The former follows from identifying the locations of the poles in $\omega$, which are fixed by the powers of $z_t$ appearing in the amplitude. A pole at $\omega_0 \in \mathbb{Z}$ contributes a factor $(i|z_t|)^{\omega_0}$ through the exponential in~\eqref{final}, yielding a real or imaginary term depending on whether $\omega_0$ is even or odd. Since $a(t,\omega)$ is real for $\omega \in \mathbb{R}$, we conclude that even (odd) powers of $z_t$ in the amplitude come with real (imaginary) coefficients.

We now turn to the powers of $L$. Any power of $L$ appearing in the discontinuity $\mathrm{Disc}_s\,\mathcal{M}_{2 \to 2}$ produces a higher–order pole in $a(t,\omega)$ from \eqref{eq:disc_a}: a term proportional to $L^r$ in the discontinuity leads to a pole of order $r+1$ in $\omega$. In the dispersion integral \eqref{final}, such a pole contributes a factor of $L^r$ to the full amplitude through derivatives acting on $e^{\omega L}$. The EFT power counting implies that the discontinuity can contain at most one logarithm per loop~\cite{Chiu:2012ir,Duhr:2014woa}. Furthermore, logarithms in $\mathrm{Disc}_s\,\mathcal{M}_{2 \to 2}$ arise only from collinear or soft corrections; inserting an additional Glauber exchange does not generate any new logs. 
So at $\ell$ loops the series will be 
\vspace{-3pt}
\begin{align}
\mathcal{M}_{2\rightarrow 2}^{(\ell)} \sim & \; c_0 z_t^{\ell+2} + z_t^{\ell+1}\left(c_{1,1}L+c_{1,0}\right) + \\
& + z_t^{\ell}(c_{2,2}L^2 +c_{2,1}L+c_{2,0}) + \nonumber \\ 
& + \cdots z_t^2 (c_{\ell,\ell}L^{\ell}+c_{\ell,\ell-1}L^{\ell-1}+ \cdots c_{\ell,0}) \,. \nonumber
\end{align}
Thus the maximal power of logarithms at $\ell$ loops is $\ell$. Furthermore, each power $z_t^j$ is accompanied by a polynomial in $L$ whose highest power is $\min(\ell+2-j,\ell)$. Combining this structure with the correlation between powers of $z_t$ and phases of $i$ obtained from~\eqref{final} leads to our final expression for the $\ell$-loop amplitude, Eq.~\eqref{eq:ampl-high-energies}.

Finally, we note that our logarithmic power counting excludes hard-region corrections. As shown in~\cite{Bartels:2012ra}, hard contributions can generate double logarithms, which we have not incorporated here. These double logs are purely quantum and are parametrically suppressed, although at sufficiently high orders in $G_N$ they may be enhanced to a (super)classical scaling. Their appearance represents yet another manifestation of the breaking of eikonalization.

As a cross-check, we can use this result to reproduce the dispersion–relation formula quoted in Appendix~A of~\cite{DiVecchia:2020ymx}, which states that if 
$\mathrm{Im}\,\mathcal{M}_{2 \to 2}(s,t) \sim s^n \log^p(s/|t|)$ at leading logarithmic accuracy, then
\begin{align}
   \hspace{-12pt} \frac{\text{Re}\,\mathcal{M}_{2 \to 2}}{\text{Im}\,\mathcal{M}_{2 \to 2}} &= -\frac{2\log(s/|t|)}{(1 + p)\pi}+ O(\log^0(s/|t|)),\quad \text{$n$ even},\nonumber \\
   \hspace{-12pt} \frac{\text{Re}\,\mathcal{M}_{2 \to 2}}{\text{Im}\,\mathcal{M}_{2 \to 2}} &= \frac{\pi p}{2\log(s/|t|)}+ O(\log^{-2}(s/|t|)),\quad \text{$n$ odd} .
    \label{Appendix A}
\end{align}
To see that Eq.~\eqref{eq:ampl-high-energies} reproduces this behaviour, consider the component of the amplitude scaling as $z_t^{\,n} L^{p}$. Using $z_t^n \propto s^n + O(t/s)$ and $\log(z_t) = \log(s/|t|) + \log(2) + O(t/s)$ we obtain, for even $n$,
\begin{align}
    \mathcal{M}_{2 \to 2}&\stackrel{\text{$n$ even}}{\sim} s^n L^{p + 1}\\
    & = s^n\left(\log^{p +1}(s/|t|) -(p + 1) \frac{i\pi}{2}\log^p(s/|t|) + \dots \right),\nonumber 
\end{align}
while for odd $n$ we find
\begin{align}
     \mathcal{M}_{2 \to 2}&\stackrel{\text{$n$ odd}}{\sim} i s^n L^{p}   \\
     &=s^n\left(i \log^{p}(s/|t|) +p \frac{\pi}{2}\log^{p-1}(s/|t|) + \dots \right), \nonumber 
\end{align}
ellipses denote terms subleading in powers of $\log(s/|t|)$. Taking ratios of the real and imaginary parts in the two cases reproduces Eq.~\eqref{Appendix A}. As a further cross-check on eq.\eqref{eq:ampl-high-energies}, we have compared it with the existing amplitudes in the literature, specifically the high energy limit of amplitude for GR and two massive scalars~\cite{KoemansCollado:2019ggb, Bjerrum-Bohr:2021din} and the $\mathcal{N} = 8$ supergravity amplitude through two loops~\cite{DiVecchia:2021bdo}, and found agreement in all cases.

\bibliography{references.bib}

\end{document}